 \documentclass[final,5p,times,twocolumn]{elsarticle}

\biboptions{sort&compress}

  \usepackage{graphicx}
  \usepackage{epstopdf}
  \usepackage{verbatim}

\usepackage{amssymb}
\usepackage{color}

\journal{Physics Letters A}

\begin{document}

\begin{frontmatter}



\title{Bragg Grating Rogue Wave}

\author[Roma]{Antonio Degasperis}
 \author[Brescia]{Stefan Wabnitz}
 \ead{stefan.wabnitz@unibs.it}
  \author[Dallas]{Alejandro B. Aceves}

\address[Roma] { Dipartimento di Fisica, ``Sapienza'' Universit\`a di Roma, P.le A. Moro 2, 00185 Roma, } \address[Brescia]{Dipartimento di Ingegneria dell'Informazione,  Universit\`a degli Studi di Brescia, via Branze 38, 25123, Brescia, Italy}
 \address[Dallas]{Southern Methodist University, Dallas, USA}

\begin{abstract}
We derive the rogue wave solution of the classical massive Thirring model, that describes nonlinear optical pulse propagation in Bragg gratings. Combining electromagnetically induced transparency with Bragg scattering four-wave mixing, may lead to extreme waves at extremely low powers.
\end{abstract}


\end{frontmatter}




\section{Introduction}
\label{sec:intro}
Extreme wave phenomena appear in a variety of scientific and social 
contexts, ranging from hydrodynamics and oceanography to geophysics, 
plasma physics, Bose-Einstein condensation (BEC), financial markets and nonlinear 
optics \cite{solli2007}-\cite{dud14}. 
Historically, the first reported manifestation of extreme or rogue
 waves is the sudden appearance in the open sea of an isolated giant wave, 
with height and steepness much larger than the average values of ocean 
waves. A universal model for describing the dynamics of rogue wave 
generation in deep water with a flat bottom 
is the one-dimensional
nonlinear Schr\"odinger (NLS) equation in the self-focusing regime. 
The mechanism leading to the appearance of NLS rogue waves requires 
nonlinear interaction and modulation instability (MI) of the continuous 
wave (CW) background \cite{benjamin}. Indeed, the nonlinear development 
of MI may be described by families of exact solutions such as the Akhmediev 
breathers \cite{akhmediev86}, which are recognized as a paradigm for rogue 
wave shaping. A special member of this solution family is the famous 
Peregrine soliton \cite{peregrine83}, which describes a wave that appears 
from nowhere and disappears without a trace.
Extreme waves that may be well represented by the Peregrine soliton have 
recently been experimentally observed in optical fibers \cite{kibler2010}, 
in water-wave tanks \cite{amin11} and in plasmas \cite{bailung11}.

Moving beyond the one-dimensional NLS model, it is important to consider 
extreme wave phenomena in either multidimensional or multicomponent 
nonlinear propagation. Vector systems are characterized by the possibility 
of observing a coupling of energy among their different degrees of freedom,
which substantially enriches the complexity of their rogue-wave families. 
Recent studies have unveiled the existence of extreme wave solutions in the
vector NLS equation or Manakov system  
\cite{baronio12}-\cite{baronio14}, 
the three-wave resonant interaction equations \cite{baronio13}, 
the coupled Hirota equations \cite{chen13} and the long-wave-short-wave 
resonance \cite{chen14R}. 

In this Letter, we present the rogue wave solution of the classical massive 
Thirring model (MTM) \cite{thir58}, a two-component nonlinear wave evolution 
model that is completely integrable by means of the inverse scattering 
transform method \cite{mik76}-\cite{kaup77}.  The classical MTM  is a 
particular case of the coupled mode equations (CMEs) that describe pulse 
propagation in periodic or Bragg nonlinear optical media 
\cite{win82}-\cite{egg97}. Furthermore, the CMEs also 
appear in other physical settings. In particular and relevant to 
rogue waves, they describe ocean waves in deep water for a periodic 
bottom \cite{ruban}. As such, 
the search for novel solution forms of these equations including rogue
waves, provides understanding of nonlinear phenomena and leads to 
applications beyond optical systems. In this respect, benefiting from   
the result \cite{ale89, ale892} that many MTM solutions (including 
single and multi-solitons and cnoidal-waves) may be mapped 
into solutions of the CMEs, provides a tool used in several works 
including  ocean waves \cite{ruban2}, BEC \cite{meystre} and metamaterials
\cite{longhi}. 

After discussing the analytical rogue wave solution in section 
(\ref{sec:theo}), in section (\ref{sec:resu}) we numerically confirm its 
stability, and show that it may also be applied to describe the generation 
of extreme events in the more general context of the CMEs. Finally, 
in section (\ref{sec:concl}) we discuss the physical implementation of 
MTM rogue waves by using coherent effects in resonant nonlinear media, 
such as electromagnetically induced transparency (EIT), which may lead to 
the giant enhancement of cross-phase modulation (XPM) with the 
simultaneous suppression of self-phase modulation (SPM).

\section{Analytical solution}
\label{sec:theo}

Let us express the MTM equations for the forward and backward waves with 
envelopes $U$ and $V$, respectively, as \newline
\begin{eqnarray}\label{mtm}
U_\xi &=&-i\nu V -\frac{i}{\nu} |V|^2 U \nonumber \\
V_\eta&=&-i\nu U -\frac{i}{\nu} |U|^2 V \;.
\end{eqnarray}
\newline
Here the light-cone coordinates $\xi\,,\,\eta$ are related to the space 
coordinate $z$ and time variable $t$ by the relations 
$\partial_\xi = \partial_t +c \partial_z$ and 
$\partial_\eta = \partial_t - c \partial_z$, where $c>0$ is the 
linear group velocity.  Even though the arbitrary real parameter $\nu$ 
can be rescaled to unity, we find it convenient to keep it for 
dimensional reasons. 
\newline
The rogue waves travel over the following CW background 
\begin{equation}\label{cw}
U_0=ae^{i\phi}\;\;,\;\; V_0=-be^{i\phi}
\end{equation}
where, with no loss of generality, the constant amplitudes $a$ and $b$ are real, and the common phase $\phi(\xi,\eta)$ is
\begin{equation}\label{phase} 
\phi = \alpha \xi + \beta \eta\;,\; \alpha=b(\frac{\nu}{a}-\frac{b}{\nu})\;,\;\beta=a(\frac{\nu}{b}-\frac{a}{\nu}) \;.
\end{equation}
Up to this point we consider the two amplitudes $a,b$ as free background 
parameters.
It can be proved that rogue wave solutions of Eqs. (\ref{mtm}) exist if and only if the two amplitudes $a,b$ satisfy the inequality 
\begin{equation}\label{condition} 
0 \, <\,ab\,<\, \nu^2 \;.
\end{equation}
By applying the Darboux method to the MTM \cite{degas14},
 one obtains the following rogue wave solution
\begin{equation}\label{rw}
U=ae^{i\phi} \frac{\mu^*}{\mu} (1-4i\frac{q_1^* q_2}{\mu^*}) \;\;,\;\;V=-be^{i\phi} \frac{\mu}{\mu^*} (1-4i\frac{q_1^* q_2}{\mu}) 
\end{equation}
with the following definitions
\begin{eqnarray}\label{qq}
q_1&=&\theta_1(1+iq)+q\theta_2\;,\;q_2=\theta_2(1-iq)+q\theta_1\;, \nonumber \\
\;q&=&\frac{a}{\chi^*}\eta+b\chi^*\xi \;.
\end{eqnarray}
and
\begin{eqnarray}\label{mup}
\mu&=&|q_1|^2 + |q_2|^2 + (i/p) (|q_1|^2 - |q_2|^2)\;, \nonumber \\
p&=&\sqrt{\frac{\nu^2}{ab}-1} >0\;.
\end{eqnarray}
In the expression (\ref{rw}), which is the analog of the Peregrine solution of the focusing NLS equation, the 
free parameters are the two real background parameters $a,b$, which are however constrained by the condition (\ref{condition}), and the two complex parameters $\theta_1,\theta_2$, while the parameter $\chi$ is given by the expression
\begin{equation}\label{chi} 
\chi= \frac{b}{\nu} (1+ip) = \frac{\nu}{a(1-ip)}\;\;.
\end{equation}
Expression (\ref{rw}) of the rogue wave solution may be simplified by fixing the reference frame of the space-time coordinates. The general solution (\ref{rw}) may then be obtained by applying to this particular solution a Lorentz transformation.

According to the last remark above, we now provide the rogue wave solution in terms of the space
$z=c(\xi-\eta)$ and time $t=(\xi+\eta)$ coordinates directly. 
By rewriting the CW phase (\ref{phase}) in these coordinates, one obtains
\begin{eqnarray}\label{Phase} 
\phi &=& k z-\omega t\;,\; k=\frac{\nu}{2c}(1-\frac{ab}{\nu^2})(\frac{b}{a}-\frac{a}{b}) \;,\nonumber \\
\;\omega&=&-\frac{\nu}{2}(1-\frac{ab}{\nu^2})(\frac{a}{b}+\frac{b}{a}) \;,
\end{eqnarray}
where $k$ is the wave number of the background CW. Setting $a=b$ means choosing the special frame of reference such that $k=0$. Note that the other possibility $a=-b$ does not satisfy the condition that $p$ is real (see (\ref{mup})).
	From a physical standpoint, the CW background solution with $a=b$ corresponds to a nonlinear wave whose frequency $\omega=-\nu(1-a^2 / \nu^2)$ 
enters deeper inside the (linear) forbidden band-gap $\omega^2<\nu^2$ 
as its intensity grows larger. 
 A linear stability analysis of the CW background solution (\ref{cw}) shows that it is modulationally unstable for perturbations with a wavenumber $k^2<4a^2/c^2$ (for details, see \cite{ale92}). Note that modulation instability gain extends all the way to arbitrarily long-scale perturbations (albeit with a vanishing gain), a condition which has been refereed to as "baseband instability", and that is closely linked with the existence condition of rogue waves in different nonlinear wave systems (e.g., the Manakov system, see Ref.\cite{baronio14}). It is also interesting to point out that, outside the range of existence of the rogue wave solution (\ref{rw}), that is for $a^2>\nu^2$, 
the background is unstable with respect to CW perturbation with a 
finite (nonzero) gain (see Ref.\cite{ale92}).

\begin{figure}[ht]
\centering
\includegraphics[width=9cm]{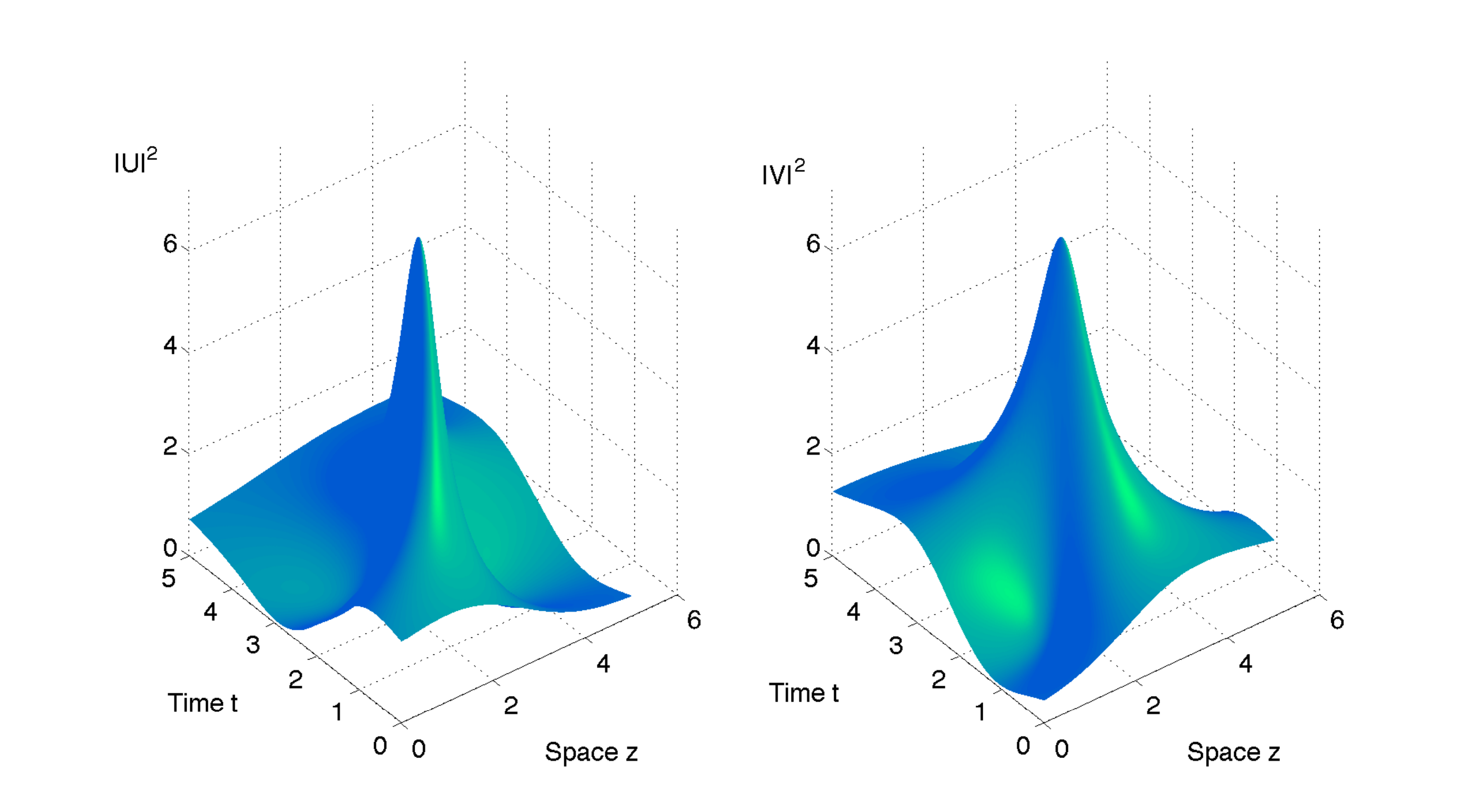} 
\caption{Evolution in time and space of the intensities in the forward and backward components of the rogue wave solution (\ref{RW}).}
\label{fig1}
\end{figure} 

By using translation invariance to eliminate the parameters $\theta_1, \theta_2$ , one finally ends up with the following expression of the MTM rogue wave solution
\begin{eqnarray}\label{RW}
U&=&ae^{-i\omega t} \frac{\mu^*}{\mu} [1-\frac{4}{\mu^*}q^* (q+i)] \;\;,\nonumber \\
\;\; V&=&-ae^{-i\omega t}  \frac{\mu}{\mu^*} [1-\frac{4}{\mu}q^* (q+i)]
\end{eqnarray}
where 
\begin{eqnarray}\label{omegaqmu}
\omega&= &-\nu (1-\frac{a^2}{\nu^2})\;,\;\;q=-\frac{a^2}{\nu c}[ip(z-z_0)-c(t-t_0)] \;,\nonumber \\ 
\;p&=&\sqrt{\frac{\nu^2}{a^2}-1}\;, \nonumber \\
\;\mu&=&2|q|^2 +(1+2\textrm{Im}q)(1-\frac{i}{p})  \;,
\end{eqnarray}
where $z_0$ and $t_0$ are arbitrary space and time shifts, respectively. Note that a further simplification may come from rescaling $z,t,U,V$ by using the length scale factor $S=-\nu c/a^2$.  

\begin{figure}[ht]
\centering
\includegraphics[width=10cm]{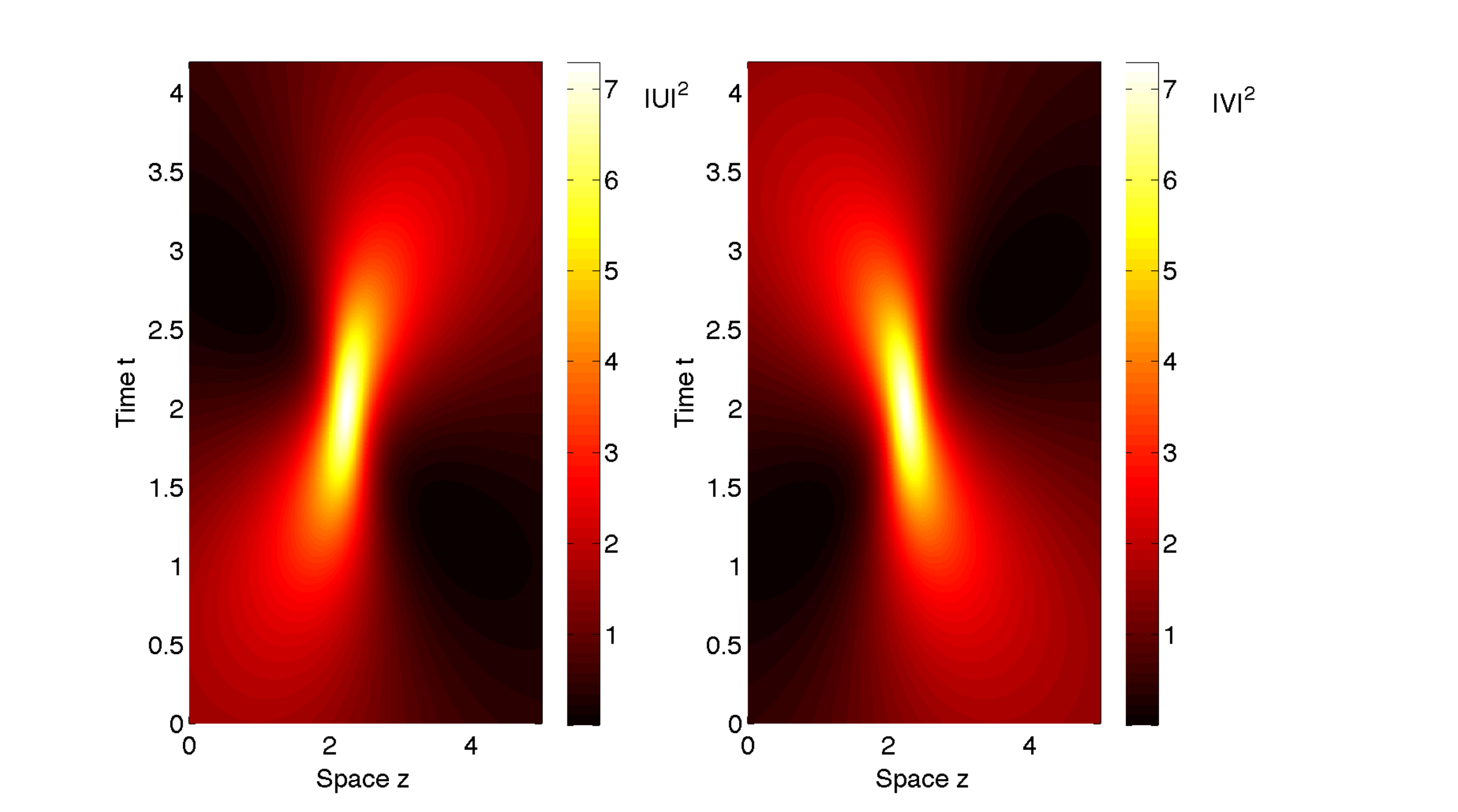} 
\caption{ Contour plot of the intensities of the forward and backward waves 
as in Fig.\ref{fig1}} 
\label{fig2}
\end{figure}

In Fig.\ref{fig1} we show the dependence on space and time of the 
intensities $|U|^2$ and $|V|^2$  of the forward and backward rogue 
waves (\ref{RW}). Here we have set $\nu=-1$, $c=1$, $a=0.9$, $t_0=2$ and 
$z_0=3.5$. As can be seen, the initial spatial modulation at $t=0$ 
evolves into an isolated peak with a maximum intensity of about 
\emph{nine times} larger than the CW background intensity. The corresponding contour plot of these intensities is shown in Fig.\ref{fig2}.

\section{Numerical results}
\label{sec:resu}

 In order to verify the spatio-temporal stability of the rogue wave solution (\ref{RW}) over a finite spatial domain,  we numerically solved Eqs. (\ref{mtm}) with $\nu=-1$, $c=1$, and using the initial (i.e., at $t=0$) and boundary (i.e., at $z=0$ and $z=L$) conditions given by the exact expression (\ref{RW}), with the same solution parameters as in Figs.\ref{fig1}-\ref{fig2}, and with $L=5\simeq4S$. 
 
\begin{figure}[ht]
\centering
\includegraphics[width=10cm]{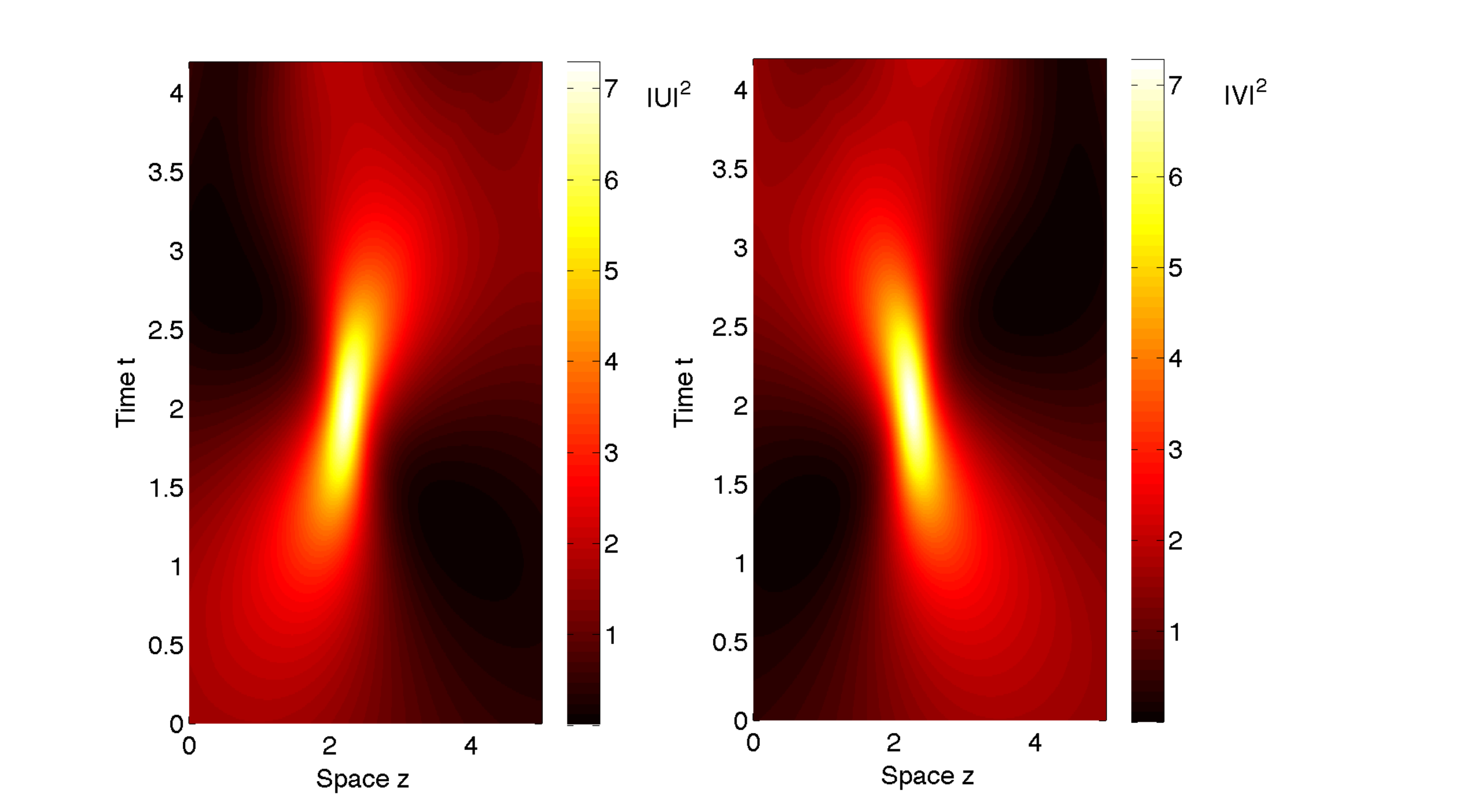} 
\caption{Numerical solution corresponding to the analytical solution in Figs.\ref{fig1}.}
\label{fig3}
\end{figure}

 Fig.\ref{fig3} displays the numerically computed intensities of the forward and backward waves: as it can be seen, there is an excellent agreement with the analytical solution. This confirms the stability and observability of the rogue wave solution (\ref{RW}), in spite of in the presence of the competing background MI. 

\begin{figure}[ht]
\centering
\includegraphics[width=10cm]{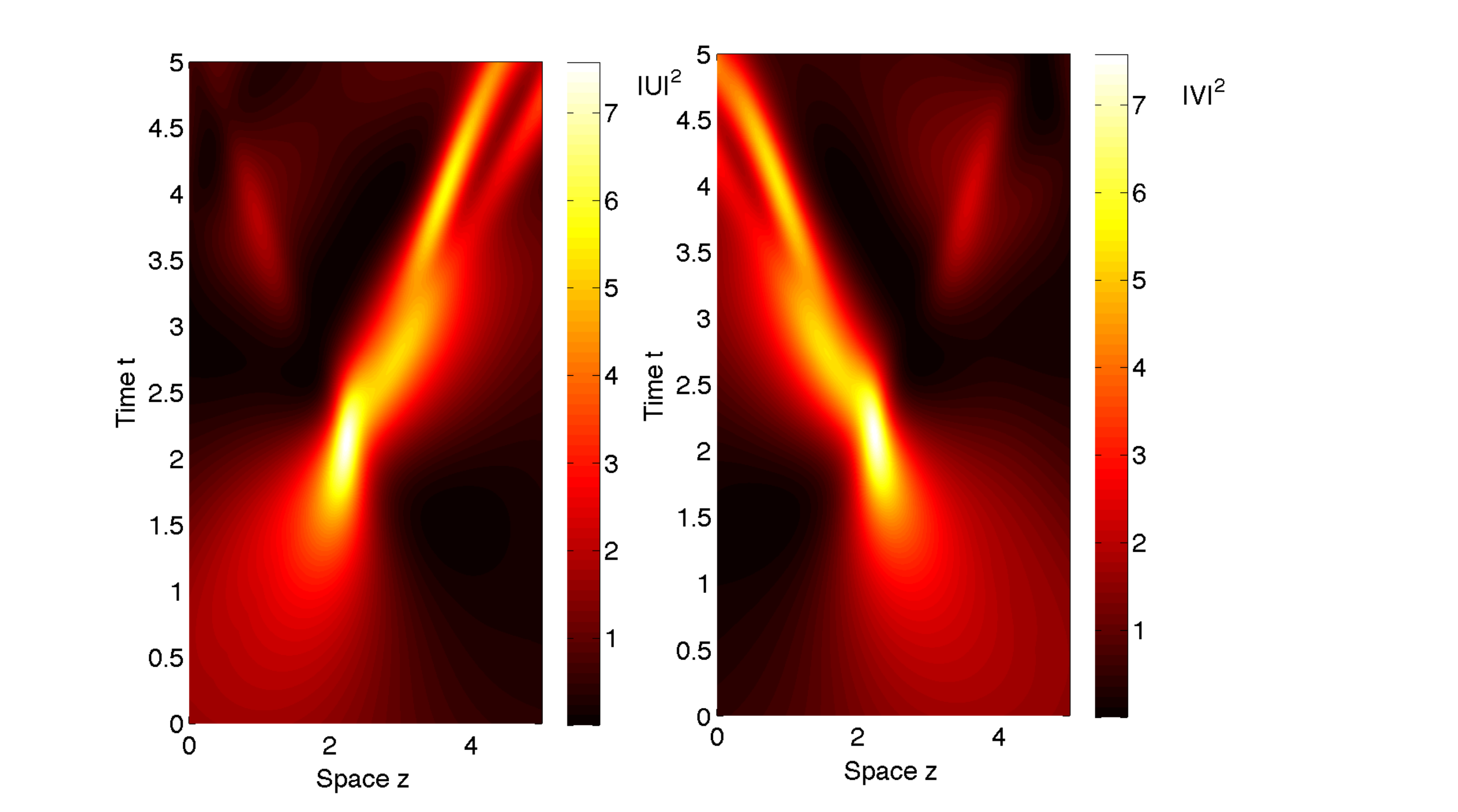} 
\caption{Numerical solution of CMEs with $\sigma=0.3$ and initial and boundary conditions as given by the analytical solution of Figs.\ref{fig1}.}
\label{fig4}
\end{figure}

\begin{figure}[ht]
\centering
\includegraphics[width=10cm]{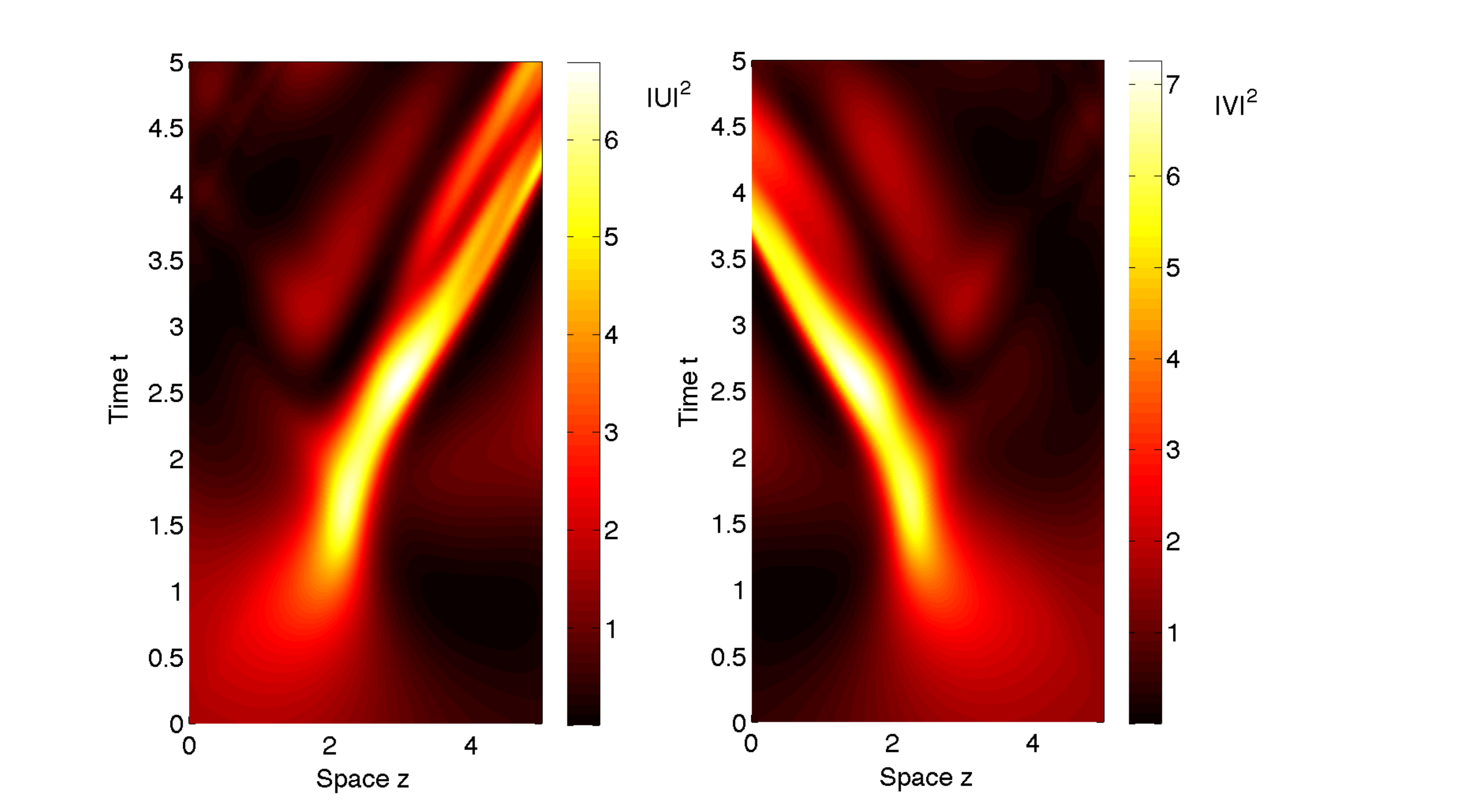} 
\caption{As in Figs.\ref{fig4}, with $\sigma=-0.5$ .}
\label{fig5}
\end{figure}

At this point it is quite natural, and interesting, to numerically check 
whether the initial conditions of the rogue wave solution
(\ref{RW}) may induce also the generation of a rogue
wave as modeled by the CMEs for pulse propagation in 
nonlinear Bragg gratings \cite{ale89}  
 or ocean waves with periodic bottom 
\cite{ruban}. Indeed, the CMEs 
\begin{eqnarray}\label{cme}
U_\xi &=&-i\nu V -\frac{i}{\nu} \left(|V|^2 +\sigma|U|^2\right) U\;,\nonumber \\
V_\eta&=&-i\nu U -\frac{i}{\nu} \left(|U|^2 +\sigma|V|^2\right) V \;.
\end{eqnarray}
differ from the MTM Eqs. (\ref{mtm}) by an additional  SPM  term, its relative strength being expressed by the coefficient $\sigma$. 
We have numerically solved Eqs.(\ref{cme}) with initial and boundary conditions as given by expression (\ref{RW}). As shown by Fig.\ref{fig4} and Fig.\ref{fig5},  quite surprisingly even in the case of a comparatively large SPM contribution, one still observes the generation of an extreme peak with nearly the same intensity and time width as in the MTM case. However now the peak no longer disappears, but it leaves behind traces in the form of dispersive waves, owing to the non-integrability nature of the CMEs. In Fig.\ref{fig4} we set $\sigma=0.3$, while qualitatively very similar results are also obtained for larger values of $\sigma$ (e.g., for $\sigma=0.5$, which corresponds to the case of nonlinear fiber gratings).

On the other hand, in Fig.\ref{fig5} we show the case with $\sigma=-0.5$, that is relevant to water wave propagation in oceans with a periodic bottom \cite{ruban}. Interestingly, in the hydrodynamic case the formation of a first rogue peak where both components have nearly the same amplitude as in the MTM case Fig.\ref{fig3}, is followed by wave breaking into secondary, and yet intense peaks in each of the two generated pulses  that
travel in opposite directions.

\section{Discussion and conclusions}
\label{sec:concl}

For the strict applicability of the analytical MTM rogue wave solution (\ref{RW}) of section (\ref{sec:theo}), it is necessary that SPM can be neglected with respect to XPM. This situation occurs whenever the forward and backward wave envelopes have different carrier frequencies, say, $\omega_U$ and $\omega_V$, and their frequency difference $\Delta\omega=\omega_U-\omega_V$ is close to a resonant frequency of the nonlinear medium \cite{akh81}. Whenever the forward and backward waves are coherently coupled in a four-level atomic system by a CW pump field, a giant enhancement of the strength of XPM is also possible via the EIT effect \cite{sch96}-\cite{fried05}, with no competing SPM. In addition, EIT brings the important benefit of zero linear absorption losses, and it is only limited by the residual two-photon absorption, which can be neglected for relatively short interaction distances. 

For maximizing the interaction length of the counter-propagating waves, diffraction effects can be suppressed by using an atomic vapor cell containing an hollow-core photonic band gap fiber (PBGF).  A scheme of Bragg soliton generation in a coherent medium exhibiting EIT has been previously discussed, however counter-propagating pulses at the same probe frequency were considered \cite{jia07}. A dynamic refractive index grating for coupling forward and backward signal waves at different frequencies may be induced by the standing wave generated from the beating of two counter-propagating pump waves, with a frequency difference equal to $\Delta\omega$. This process is known as Bragg scattering four-wave mixing (BS-FWM) \cite{mc05}, and it enables noise-free frequency translation. Indeed, BS-FWM  in the co-propagation geometry and at microwatt pump power levels has been recently demonstrated in Rubidium vapor, confined to a few centimeters long PBGF \cite{gaeta14}. 

Note that the solution (\ref{RW}) may also apply to describe extreme wave emergence in the co-propagation of two signals at different carrier frequencies \cite{wab89}, by simply interchanging the role of time and space variables in Eqs.(\ref{mtm}). In this case, a dynamic grating may be induced by the beating of two pump waves in the orthogonal polarization \cite{sima04}. 

In short summary, we obtained the rogue wave solution of the classical MTM, which extends the Peregrine soliton solution of the NLS equation to the case of wave propagation in a periodic nonlinear medium. An implementation is proposed using coherent resonant wave mixing: we envisage that Bragg solitons and rogue waves may ultimately be observed at sub-milliwatt pumping levels by using chip-scale waveguides that are evanescently coupled to atomic vapors \cite{agha13} . 

\section{Acknowledgments}
The present research was supported by Fondazione Cariplo, grant n.2011-0395, and the Italian Ministry of University and Research (MIUR) (grant contract 2012BFNWZ2).











%
%

\end{document}